\documentclass[a4paper]{jpconf}
\usepackage{graphicx}
\begin{document}
\title{SHAPIRO AND PARAMETRIC RESONANCES IN COUPLED JOSEPHSON JUNCTIONS}

\author{Ma A Gaafar$^{1,3}$, Yu M Shukrinov$^{1,2}$ and A Foda$^3$}
\address{$ˆ1$ BLTP, JINR, Dubna, Moscow Region, 141980, Russia}
\address{$ˆ2$ Max-Planck-Institute for the Physics of Complex Systems, 01187 Dresden, Germany}
\address{$ˆ3$ Faculty of Physics, Philipps-Universit$\ddot{a}$t Marburg, 35032 Marburg, Germany}
\address{$ˆ4$ Nile University, Smart Village, Cairo, Egypt}
\ead{mgaafar@theor.jinr.ru}

\begin{abstract}
The effect of microwave irradiation on the phase dynamics of intrinsic Josephson junctions in high temperature superconductors is investigated. We compare the current-voltage characteristics for a stack of coupled Josephson junctions under external irradiation calculated in the framework of CCJJ and CCJJ+DC models.

\end{abstract}

\section{Introduction}
Since the discovery of the intrinsic Josephson effect in high temperature superconductors (HTSC) such as $Bi_2Sr_2CaCu_2O_y$ (BSCCO), numerous studies of the basis nature, nonlinear dynamics and device applications of the superconductors have been extensively developed. The high-frequency response of Intrinsic Josephson junctions (IJJs) have been constantly updated. One of the most spectacular indications of the Josephson effect is locking of Josephson oscillations to the frequency of external microwave (MW) irradiation. As the Josephson frequency is proportional to voltage, the locking leads to appearance of steps at quantized voltages, called Shapiro steps \cite{Shapiro, Tinkham}. The steps appear in the current-voltage characteristic (CVC) at discrete voltage values $V_n=nh\omega/2e$ ($\omega$, the frequency of the applied MW signal; $h$, the Planck constant; $e$, the elementary charge; $n$, integer number).

It was shown by Koyama and Tachiki \cite{koyama96} that the system of equations for capacitively coupled
Josephson junctions has a solution corresponding to the longitudinal plasma wave (LPW) propagating along the c-axis. So, the Josephson oscillations can excite the LPW by their periodical actions. The frequency of Josephson oscillations $\omega_J$ is determined by the voltage value in the junction, and at
$\omega_J=2 \omega_{LPW}$, where $\omega_{LPW}$ is LPW frequency, the parametric resonance (PR) is realized. External radiation essentially
changes the physical picture of the coupled JJs. As known, the one-dimensional models with coupling between junctions capture the main features
of real IJJs, like hysteresis and branching of the CVC, and help to understand their physics. So it is of interest to investigate this system under MW irradiation using these models.

In this article we investigate the phase dynamics of the coupled Josephson junctions under MW irradiation in the framework of both CCJJ and CCJJ+DC models.

\section{Model and Method of Calculations}
Because of the thickness of the S-layers in HTSC is extremely small and comparable to the Debye length $r_D$ of the electric charge screening, therefore, there is no complete screening in a separate S-layer. It leads to the generalized scalar potential $\Phi_l$ of S-layer. The $\Phi_l$ is expressed through the electric
scalar potential $\phi_l$ and the derivative of order parameter's phase $\dot{\theta_l}$ of S-layer $l$ by $\Phi_l{(t)}=\phi_l-V_0 \dot{\theta_l}$.

Without screening of the charge in S-layer the Josephson equation is generalized $\dot{\varphi_l}=V_l+ \alpha(2V_l-V_{l-1}-V_{l+1})$ (where $\alpha$ is the coupling between junctions) and together with the expression for total current through the stack $I =\dot{V_l} +\sin\varphi_l +\beta V_l$ describe the phase dynamics of the coupled system of JJs in the framework of CCJJ model \cite{koyama96}. In CCJJ-model, the system of equations for the gauge-invariant phase differences $\varphi_l(\tau)$  between S-layers $l$ and $l+1$ in the presence of MW irradiation is described by:

\begin{eqnarray}
\frac{d^2}{d\tau^2}\varphi_{l}=(I-\sin \varphi_{l} -\beta\frac{d\varphi_{l}}{d\tau})+ \alpha (\sin \varphi_{l+1}+
\sin\varphi_{l-1} - 2\sin\varphi_{l})- A\sin\omega \tau
\end{eqnarray}
 where $I$ is the external bias current, $A$ is radiation amplitude, $\omega$ is MW radiation frequency and $\beta$ is the dissipation parameter ($\beta$ related to the McCumber parameter $\beta_c$  by $\beta_c= 1/\beta^2$). The CVC in CCJJ model are characterized by branching at $I=I_c$ (where $I_c$ is the Josephson critical current) and strong branching in the hysteresis region: the number of branches in general case is equal to $2^{N}$, where $N$ is the number of junctions in the stack.

In CCJJ+DC model \cite{seidel} the diffusion current $J^{l}_D = \Phi_{l+1}-\Phi_l$ is added and the expression for current through the stack of junctions in this model is
$I =\dot{V_l} +\sin\varphi_l +\beta \varphi_l$. Second order differential equation for $\varphi_l$ has additional terms in compare with CCJJ model which are proportional to the product of coupling and dissipation parameters
\begin{eqnarray}
\frac{d^2}{d\tau^2}\varphi_{l}=(I-\sin \varphi_{l} -\beta\frac{d\varphi_{l}}{d\tau})+ \alpha (\sin \varphi_{l+1}+
\sin\varphi_{l-1} - 2\sin\varphi_{l})
\nonumber \\+ \alpha \beta(\frac{d\varphi_{l+1}}{d\tau}+\frac{d\varphi_{l-1}}{d\tau}-2\frac{d\varphi_{l}}{d\tau}) - A\sin\omega \tau
\end{eqnarray}
To find the CVC of the stack we solve the system of dynamical equations for phase differences using the fourth order Runge-Kutta method. We use a dimensionless time $\tau = t\omega_p$, where $\omega_{p}$ is the plasma frequency $\omega_{p}=\sqrt{2eI_c/\hbar
C}$ and $C$ is the capacitance. In our simulations we measure the voltage in units of $V_0=\hbar\omega_p/(2e)$, the frequency in units of $\omega_{p}$, the current and the amplitude of radiation in units of $I_c$. To study the time dependence of the charge in the S-layers, we use the Maxwell equation
$\varepsilon\varepsilon_0\nabla. \vec{E} = $Q$ $, where $\varepsilon$ and
$\varepsilon_0$ are relative dielectric and electric constants. The charge density $Q_l$ in the S-layer $l$ is proportional to the difference between the voltages $V_{l}$ and $V_{l+1}$ in the neighbor insulating layers $Q_l=Q_0 \alpha (V_{l+1}-V_{l})$,
where $Q_0 = \varepsilon \varepsilon _0 V_0/r_D^2$.

We solve the system of dynamical equations for phase differences at fixed value of bias current $I$ in some time interval $(0, T_m)$ with the time step $\delta \tau$. This interval is used for time averaging procedure. Then we change the bias current by
$\delta I$, and repeat the same
procedure for the current $I+\delta I$ in new time interval $(T_m, 2T_m)$.
The values of the phase and its time derivative in the end of the first time interval are used as initial conditions for second time
interval and so on. The total
recorded time is calculated as $t_r= t\omega_p +T_m(I_0 -I)/\delta I$, where $I_0$ is an initial value of the bias current for time dependence recording. In our simulations we put $T_m=1000$, $\delta \tau=0.05$, $\delta I=0.0001$ and we add to the bias current a small noise with amplitude in the interval $(+10^{-8}, -10^{-8})$. The details concerning the numerical procedure are given in Ref. \cite{smp-prb07}.
\section{Results and Discussion}

In our present work, we describe the case $\omega > 2\omega_{LPW}$, when the SS is above PRR in CVC. It was shown in Ref. \cite{Gaafar2012} that an increase of the amplitude of irradiation leads to the appearance of an additional PR above SS called as radiation related parametric resonance (rrPR).

\begin{figure}[!ht]
 \centering
\includegraphics[height=61mm]{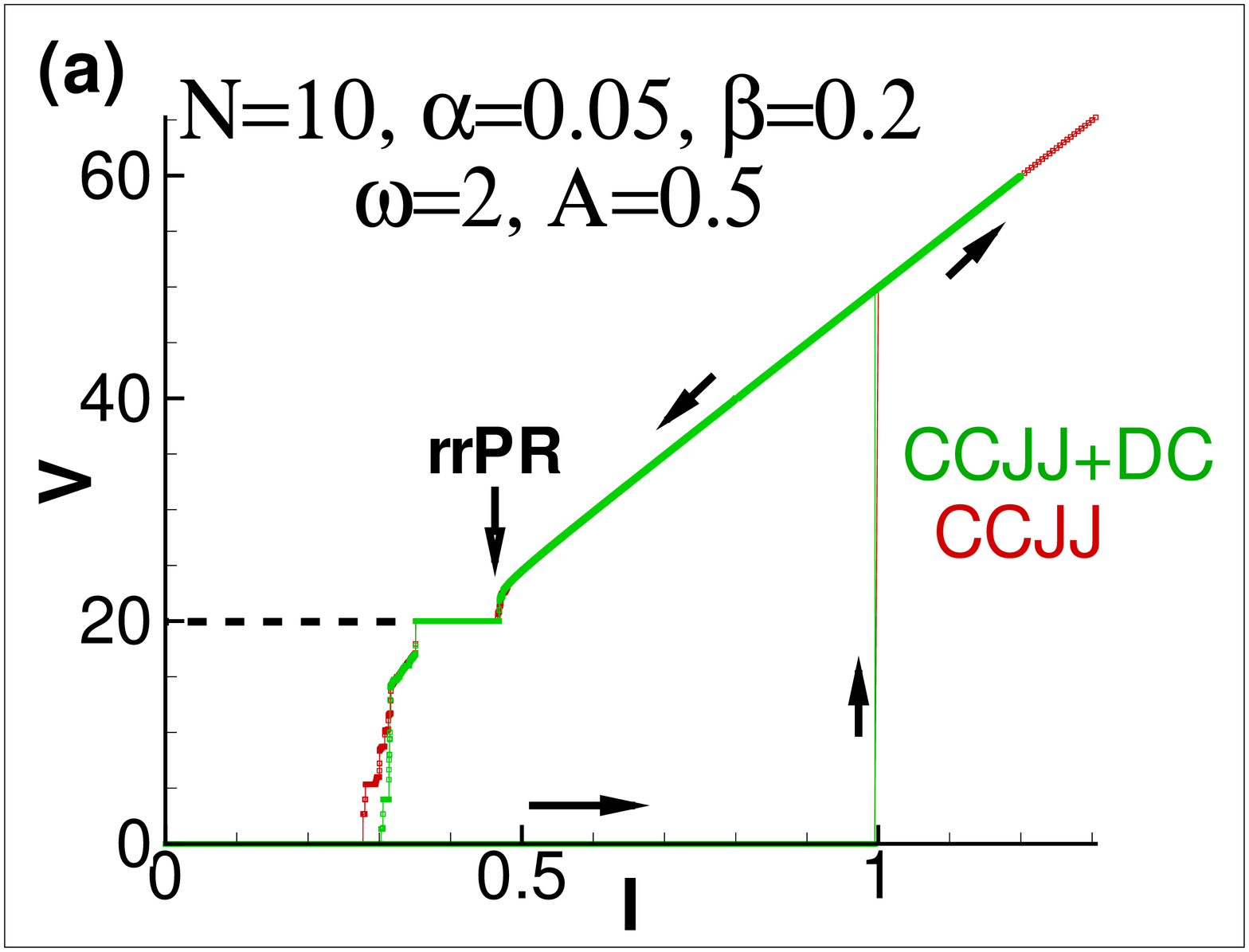}
\includegraphics[height=60mm]{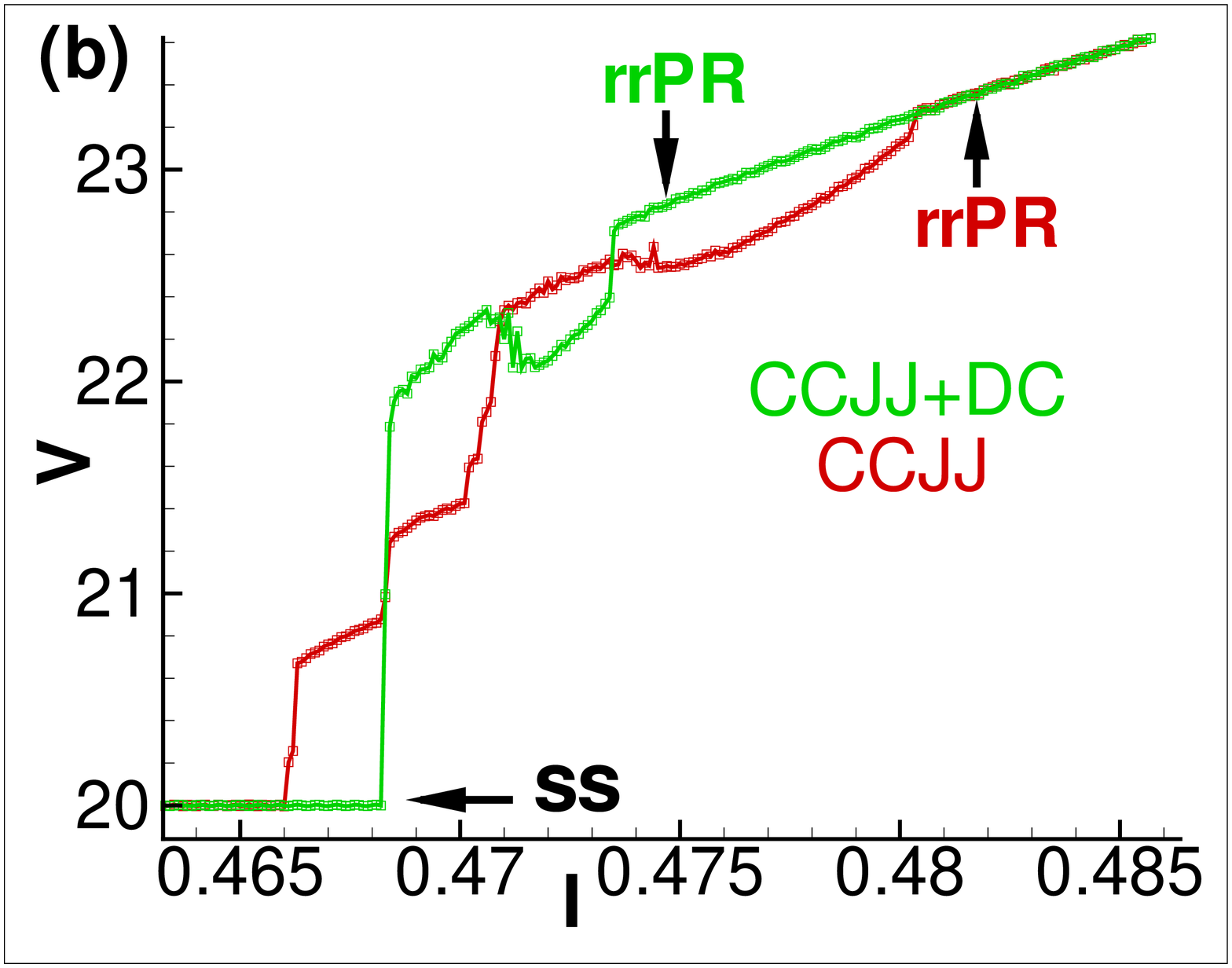}
\caption{CVC of the stack with 10 coupled JJs under microwave irradiation with $\omega=2$ and $A=0.5$ simulated in
the framework of the CCJJ and CCJJ+DC models at $\alpha=0.05$, $\beta=0.2$ and periodic boundary conditions; (b) The enlarged part of the rrPR above SS on the outermost branch of CVC.}
 \label{1}
\end{figure}

In Fig.~\ref{1}(a) we show the CVC of the stack with 10 coupled JJs under microwave irradiation with $\omega=2$ and $A=0.5$ simulated in the framework of both the CCJJ and CCJJ+DC models at $\alpha=0.05$, $\beta=0.2$ and periodic boundary conditions. At $\omega=0$, PR is characterized by breakpoint current $I_{bp}\simeq 0.28$, and breakpoint voltage $V_{bp} \simeq 11.5$, corresponding to the Josephson frequency $\omega_J=1.15$.

As we can see, the first SS is developed on the outermost branch of CVC at $V=\omega_J*N=20$. Dashed line stresses this fact. Both models are coincide with the increase in bias current. The difference between CCJJ and CCJJ+DC models appears in the branching part of CVC in hysteretic region. The CCJJ model demonstrates more intensive branching and larger hysteresis region in compare with CCJJ+DC model. This feature could be explained by the role of diffusion current \cite{Dubna2010}.

\begin{figure}[!ht]
 \centering
\includegraphics[height=60mm]{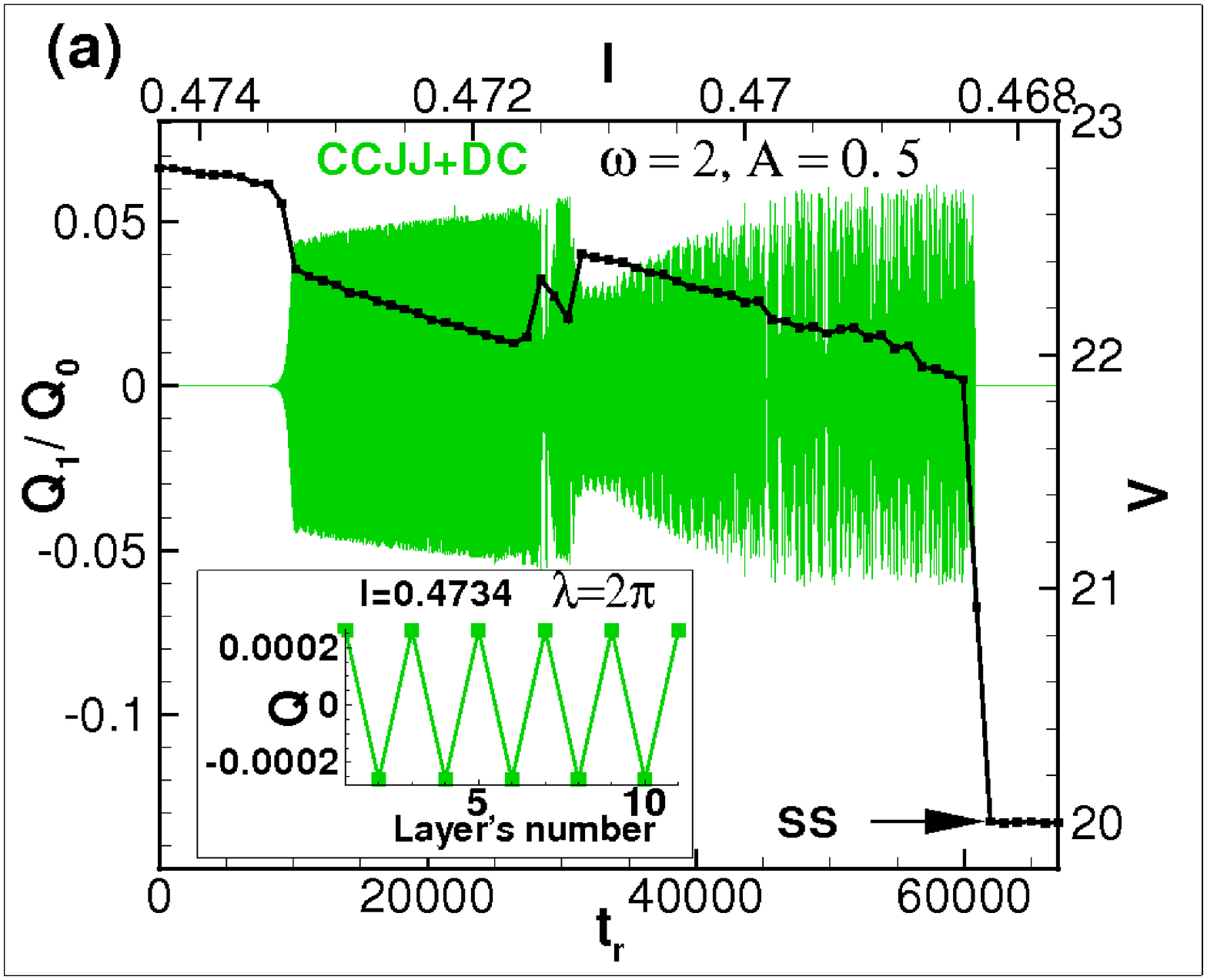}
\includegraphics[height=60mm]{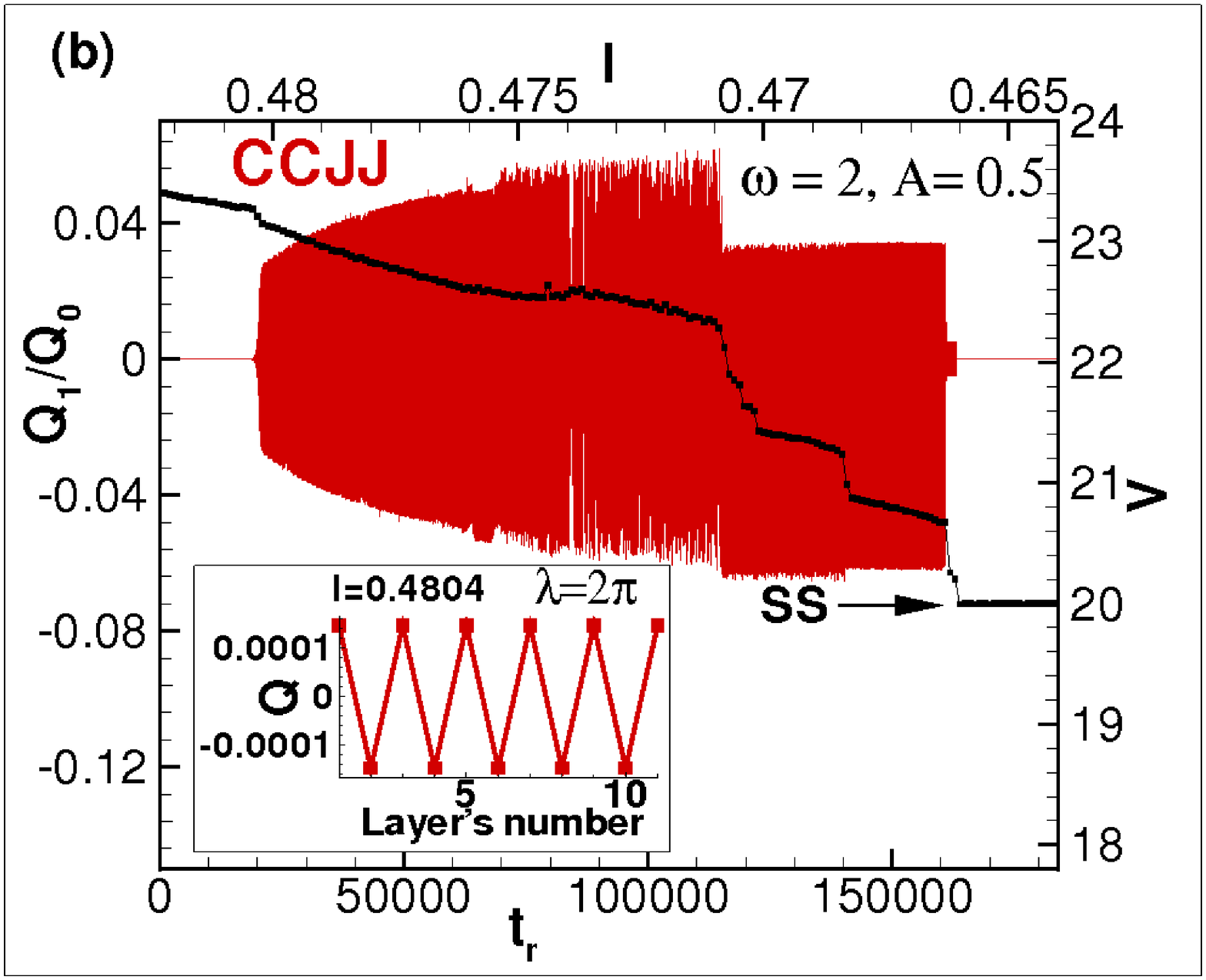}
\caption{Charge time-dependence in the radiation related parametric resonance region of the stack with ten coupled JJs at $\alpha= 0.05$, $\beta= 0.2$, $\omega=2$ and $A=0.5$ calculated in the framework of (a) CCJJ+DC-model; (b) CCJJ-model. The insets show the charge distribution among the S-layers in the growing region.}
 \label{2}
\end{figure}

As we mentioned previously, an increase of the amplitude of irradiation leads to the appearance of rrPR. Fig.~\ref{1}(b) shows (for both models) an enlarged view of the rrPR region on the outermost branch of the CVC (the temporal oscillations of the charge in this region are shown in Fig. ~\ref{2}). We see that rrPR is started at higher currents for CCJJ than CCJJ+DC model. In addition, we observe an interesting structure before SS in CCJJ-model.

Fig.~\ref{2}(a) demonstrates the temporal oscillations of the charge in the growing region of rrPR for the first S-layer of the stack with ten coupled JJs at $\alpha= 0.05$, $\beta= 0.2$, $\omega=2$ and $A=0.5$, combined with CVC of the outermost branch, calculated in the framework of CCJJ+DC model. The filled squares mark the bias current steps in CVC.
We see in this figure the correlation between the charge-time dependence and  CVC. The irradiation change the character of the charge-time dependence essentially and bring about a "bump" structure on the outermost branch of CVC, as shown in the figure. Analogies features in CVC were recently observed experimentally \cite{Benseman}. The inset shows the charge distribution among the layers in the growing region. The charge on the neighbor layers is equal in magnitude and opposite in sign. This is true for all adjacent layers and corresponds to the $\pi$-mode i. e. the wavelength of the created LPW is equal to $\lambda=2 d$.

The time dependence of the charge in the rrPR region calculated in the framework of CCJJ-model is presented in Fig.~\ref{2}(b). We see that the charge on the S-layer appears at a higher value of current ($I_{bp}=0.4809$) in comparison with the current value in case of CCJJ+DC-model ($I_{bp}=0.4739$). The inset shows the charge distribution among the layers in the growing region, illustrates that the wavelength of LPW is the same as for CCJJ+DC-model ($\lambda=2 d$).

\section{Summary}
We presented the CVC of the stack with 10 coupled Josephson junctions under microwave irradiation calculated in the framework of CCJJ and CCJJ+DC models. The CVC demonstrate different features in the radiation related parametric resonance region above SS. These results stress the essential role of the diffusion current in formation of the CVC of intrinsic Josephson junctions in high temperature superconductors. We showed that the radiation might bring to the "bump" structures
in CVCs recently observed experimentally.
\section{Acknowledgment}
We would like to acknowledge the help received through the agreement between Egypt and Joint Institute for Nuclear Research, Dubna, Russia. M. Gaafar thanks T. Hussein and H. Elsamman.

\end{document}